\documentclass{ws-ijmpa-blank-new}

\usepackage[super,compress]{cite}
\usepackage{graphicx}

\usepackage{amssymb,epsf,latexsym,amssymb,amsmath,mathrsfs}
\usepackage[english]{babel}

\newcommand{\beq}{\begin{equation}}
\newcommand{\eeq}{\end{equation}}
\newcommand{\beqa}{\begin{eqnarray}}
\newcommand{\eeqa}{\end{eqnarray}}
\newcommand{\bsubeqs}{\begin{subequations}}
\newcommand{\esubeqs}{\end{subequations}}
\newcommand{\half}{{\textstyle \frac{1}{2}}}

\makeatletter
\@addtoreset{equation}{section}
\renewcommand{\theequation}{\thesection.\arabic{equation}}
\makeatother

\begin{document}
\markboth{F.R. Klinkhamer, G.E. Volovik}
{Propagating $q$-field and $q$-ball solution}

%
\catchline{}{}{}{}{}
%

\title{\vspace*{-11mm}
Propagating $q$-field and $q$-ball solution}

\author{F.R. Klinkhamer}

\address{Institute for Theoretical Physics, Karlsruhe Institute of
Technology (KIT),\\ 76128 Karlsruhe, Germany\\
frans.klinkhamer@kit.edu}

\author{G.E. Volovik}

\address{Low Temperature Laboratory, Department of Applied Physics,
Aalto University,\\
PO Box 15100, FI-00076 Aalto, Finland,\\
and\\
Landau Institute for Theoretical Physics,
Russian Academy of Sciences,\\
Kosygina 2, 119334 Moscow, Russia\\
volovik@ltl.tkk.fi}

\maketitle


\begin{abstract}
One possible solution of the cosmological constant problem
involves a so-called $q$-field, which self-adjusts so as
to give a vanishing gravitating vacuum energy density
(cosmological constant)
in equilibrium. We show that this $q$-field can manifest
itself in other ways. Specifically, we establish
a propagating mode ($q$-wave) in the nontrivial vacuum and find
a particular soliton-type solution in flat spacetime,
which we call a $q$-ball by analogy with the well-known
$Q$-ball solution.
Both $q$-waves and $q$-balls are expected to play a role
for the equilibration of the $q$-field in the very early universe.
\end{abstract}
\vspace*{.0\baselineskip}
{\footnotesize
\vspace*{.25\baselineskip}
\noindent \hspace*{5mm}
\emph{Journal}: \emph{Mod. Phys. Lett. A} \textbf{32} (2017) 1750103
\vspace*{.25\baselineskip}
\newline
\hspace*{5mm}
\emph{Preprint}: arXiv:1609.03533 
}
\vspace*{-5mm}\newline
\ccode{PACS Nos.: 03.65.Pm, 05.45.Yv, 95.36.+x, 98.80.Es}


\section{Introduction}
\label{sec:Introduction}

A novel approach to the cosmological constant problem,
guided by thermodynamics and Lorentz invariance,
is provided by the $q$-theory
of the quantum vacuum~\cite{KlinkhamerVolovik2008a}.
The variable $q$ describes the phenomenology of the quantum vacuum.
Various realizations of $q$ have been discussed in Refs.~\citen{KlinkhamerVolovik2008a,KlinkhamerVolovik2008b,KlinkhamerVolovik2008c,%
KlinkhamerVolovik2009,KlinkhamerVolovik2010,KlinkhamerVolovik2016a,%
KlinkhamerVolovik2016b}.
Here, we only
consider the four-form-field-strength realization based on
a three-form gauge field~\cite{DuffNieuwenhuizen1980,Aurilia-etal1980,%
Hawking1984,Duff1989,DuncanJensen1989,BoussoPolchinski2000,%
Aurilia-etal2004,Wu2008}.

In Refs.~\citen{KlinkhamerVolovik2008b,KlinkhamerVolovik2008c,%
KlinkhamerVolovik2016a}, we have considered $q$-theory applied
to the behavior of the quantum vacuum in an expanding universe. Now, we concentrate on the properties of the $q$-field itself and
address the following two questions:
does the $q$-field have a propagating degree of freedom in the nontrivial vacuum and are there soliton-type solutions for the $q$-field?
The present article gives an affirmative answer to both questions.
Throughout, we use natural units with $c=\hbar=1$
and take the metric signature $(-+++)$.

\section{Kinetic term}
\label{sec:Kinetic-term}

For the $q$-theory approach to the cosmological constant problem,
it suffices to consider only a potential term of the
(pseudo-)scalar composite field $q$ in the low-energy
effective action. But nothing excludes having a
kinetic term of the composite $q$ in the effective action. In fact,
the $q$-field effective action with kinetic term can be chosen
as follows:
\bsubeqs\label{eq:actionF-Fdefinition12}
\beqa
\hspace*{-10mm}
S&=&- \int_{\mathbb{R}^4}
\,d^4x\, \sqrt{-g}\,\left[\frac{R}{16\pi G(q)} +\epsilon(q)
+\frac{1}{8}\,K(q)\, g^{\alpha\beta}\, \nabla_\alpha (q^2) \nabla_\beta (q^2) \right] \,,
\label{eq:actionF}\\[2mm]
\hspace*{-10mm}
q^2 &\equiv&- \frac{1}{24}\,
F_{\alpha\beta\gamma\delta}\,F^{\alpha\beta\gamma\delta}\,,\quad
F_{\alpha\beta\gamma\delta}\equiv
\nabla_{[\alpha}A_{\beta\gamma\delta]}\,,
\label{eq:Fdefinition1}
\\[2mm]
\hspace*{-10mm}
F_{\alpha\beta\gamma\delta}&=&q\sqrt{-g} \,\epsilon_{\alpha\beta\gamma\delta}\,,
\quad
F^{\alpha\beta\gamma\delta}=q \,\epsilon^{\alpha\beta\gamma\delta}/\sqrt{-g}\,, \quad
\label{eq:Fdefinition2} \eeqa
\esubeqs
where the functions $ G(q)$, $\epsilon(q)$, and $K(q)$
in \eqref{eq:actionF}
involve only even powers of $q$, as $q$
is a pseudoscalar according to \eqref{eq:Fdefinition2}
with the Levi--Civita symbol $\epsilon_{\alpha\beta\gamma\delta}$.
In this realization, $q$ has mass dimension 2.
A further term for the action \eqref{eq:actionF} will be presented
in Sec.~\ref{subsec:Surface-term}.

Variation of the action \eqref{eq:actionF} over the three-form
gauge field $A$
gives a generalized Maxwell equation, which has been derived in
Ref.~\citen{KlinkhamerVolovik2008b} for the theory
without direct $q$-kinetic term ($K=0$).
The presence of the $K$ term in the action \eqref{eq:actionF}
does not change the general form of the field equation,
\begin{equation}\label{eq:MaxwellGeneralized}
\nabla_\beta \left( \frac{\delta S}{\delta q}
\right) =0\,,
\end{equation}
which gives for the case at hand
\beqa\label{eq:Maxwell}
&&
\nabla_\beta \Bigg(
\frac{d\epsilon(q)}{d q}+\frac{R}{16\pi} \frac{dG^{-1}(q)}{d q}
+
\frac{1}{8}\, \frac{dK(q)}{dq}\, \nabla_\alpha (q^2) \nabla^\alpha (q^2)
-
\frac{1}{2}\,
q\, \nabla^\alpha \Big[K(q)\,\nabla_\alpha  (q^2)\Big]
\Bigg)
\nonumber\\[1mm]
&&
=0\,.
\eeqa
The $K$ term of \eqref{eq:actionF} also gives rise to an extra contribution
in the generalized Einstein equation, which need not concern us for the
moment.

The generalized Maxwell equation \eqref{eq:Maxwell}
has two constant solutions in flat Minkowski spacetime:
\begin{enumerate}
\item
the trivial solution $q=0$, which corresponds
to the ``absolutely empty vacuum'';\vspace*{1mm}
\item
 the nontrivial solution $q=q_{0}\ne 0$ (with an integration constant $\mu=\mu_{0}$), which corresponds to the ``physical vacuum.''
\end{enumerate}
The last type of constant solution can be explained as follows.
For the homogeneous vacuum, we have $d\epsilon(q)/d q=\mu$,
which determines $q$ for a fixed integration constant $\mu$.
Hence, $\mu$ plays the role
of a chemical potential, which is thermodynamically conjugate to the density $q$ of the conserved total charge $Q$.
This emergence of a conservation law makes the four-form field $F$
one of the possible realizations of the vacuum field $q$.
In the Minkowski vacuum, the equilibrium value
$\mu=\mu_{0}$ and the corresponding $q$-field value
$q_{0}\ne 0$  naturally give
a zero value for the gravitating vacuum energy density (effective cosmological constant)~\cite{KlinkhamerVolovik2008a,KlinkhamerVolovik2008b}.

The solution $q=0$, on the other hand, may be considered as
the ``absolutely empty vacuum,'' that is, a vacuum which is devoid of any type of quantum field and of the corresponding quantum fluctuations.
In the condensed matter analogy,
where the role of the variable $q$ is played
by the number density $n$ of atoms,
this vacuum corresponds to the state with $n=0$,
i.e., to empty space devoid of atoms
and, thus, of any type of emergent quantum field.

The aim of the present article is to look for nonconstant solutions
of the $q$-field.

\section{Propagating mode}
\label{sec:Propagating-mode}

We now make the linear expansion
\beq\label{eq:q-field-scalar-mode}
q(x)=q_{0} \,\big[1- \varphi(x)\big]
\eeq
and consider the integrated Maxwell equation \eqref{eq:Maxwell}
in the Minkowski spacetime background with integration constant $\mu$
fixed to the equilibrium value $\mu_{0}$.
The effect of $\varphi$ on the metric $g_{\mu\nu}$ in the generalized
Einstein equation is quadratic in $\varphi$ and can be neglected in linear approximation.  As a result, we obtain
the linear equation for the mode $\varphi$ of the $q$-field from the four-form field $F$ propagating in the Minkowski background.
This equation corresponds to the Klein--Gordon equation of a massive
scalar field.

The expression for the mass-square of the
$\varphi$-mode can be also obtained from
(\ref{eq:actionF}) by expanding in $\varphi$
and considering the quadratic terms,
 \begin{equation}
 S=- \int_{\mathbb{R}^4}
\,d^4x\,  \left( \frac{1}{2\chi_{0}}\, \varphi^2
+  \frac{1}{2}\,q_{0}^4 \,K(q_{0})\, \nabla_\alpha\varphi  \nabla^\alpha\varphi + \cdots \right) \,,
\label{eq:actionPhi}
\end{equation}
where  $\chi_{0}$  is the compressibility of the vacuum introduced in Ref.~\citen{KlinkhamerVolovik2008a},
\begin{equation}
\big(\chi_{0}\big)^{-1} \equiv
\Bigg[\, q^2\;\frac{d^2  \epsilon(q)}{d  q^2}\,\Bigg]_{q=q_{0}} > 0\,.
\label{eq:chi_{0}}
\end{equation}
From \eqref{eq:actionPhi}, the mass-square of the
$\varphi$-mode is then given by
\begin{equation}
M^2 =\frac{1}{\chi_{0}\, K(q_{0})\, q_{0}^4} \,.
\label{eq:M-square}
\end{equation}
In the absence of the kinetic term, that is for $K=0$,
the mass $M$ becomes infinite and the $q$-wave does not propagate.
Note also that the equivalence between the $q$-field
from the four-form field $F$
and the scalar field $\varphi$ exists only at the perturbative level.
In general, this $q$-field
is a composite pseudoscalar field,
with different properties compared to those of a fundamental pseudoscalar field  (see, e.g., Sec. 2 of Ref.~\citen{KlinkhamerVolovik2016b}).

If the $q$-field is of the Planck scale
[$q_0 \sim (E_P)^2 \equiv 1/G_N$, with Newton's gravitational constant
$G_N$], then the mass $M$ is also of the Planck scale,
$M^2  \sim  G_N  / \chi_{0} \sim 1/G_N = (E_P)^2$.
In the absolutely empty vacuum with $q=0$ replacing $q_{0}\ne 0$,
the vacuum $q$-field from the four-form field $F$ is not propagating
according to \eqref{eq:M-square},
which corresponds to the standard result for the
three-form gauge field $A$~\cite{DuffNieuwenhuizen1980,Aurilia-etal1980}.

It appears that there is an important difference between the ground state of condensed matter and the relativistic vacuum.
In condensed matter physics, the compressibility of the ground state determines the speed of sound in the
medium (cf. Chap. VIII in Ref.~\citen{LandauLifshitz1959}).
In the relativistic vacuum, this is impossible due to
the Lorentz invariance of the vacuum,
which selects the Lorentz-invariant form for the spectrum of propagating modes, $E^2=M^2+c^2\,p^2$.
As $c$ is fixed by the  Lorentz invariance,
only the vacuum compressibility may enter the mass $M$ of the vacuum mode,
just as we have obtained in \eqref{eq:M-square}
for the particular case of the $4$-form realization of the vacuum $q$-field.
The same should be valid in the
general case: for any realization of the vacuum field $q$,
the mass of the propagating mode is determined by the vacuum compressibility.

At this moment, we have not much to say about
the generation of these $q$-waves.
For the action \eqref{eq:actionF} with $q$-independent
standard-model fields added~\cite{KlinkhamerVolovik2016a},
the emission of $q$-waves
requires strong gravitational fields,
which may occur in the very early universe.

\section{Heuristics of the $q$-ball}
\label{sec:Heuristics}

The  $q$-ball solution of $q$-theory is a ball with radius $R$
having $q\ne 0$ inside ($r<R$) and
the empty-vacuum value $q=0$ outside ($r\geq R$).

Generally speaking, $Q$-balls are nontopological solitons
which carry a conserved global quantum number $Q$
(see Refs.~\citen{Rosen1968a,Rosen1968b,FriedbergLeeSirlin1976,Coleman1985,%
Kusenko1997,LeePang1992} and references therein).
The dynamics of the $q$-ball of $q$-theory is, however, rather different
from that of the standard $Q$-ball in, for example,
the $SO(2)$-invariant theory of two real scalar fields $\phi_1$
and $\phi_2$ with nonderivative interactions~\cite{Coleman1985}.
For this reason, we employ a somewhat different
notation, `$q$-ball' instead of `$Q$-ball.'

Incidentally, we could also consider an extended $q$-theory
with two three-form gauge fields ($A_1$ and $A_2$) and an
$SO(2)$-invariant action in terms
of the corresponding composite pseudoscalars $q_1$ and $q_2$
replacing the fundamental scalars $\phi_1$ and $\phi_2$.
In such an extended $q$-theory, there would be a direct
analog of the standard $Q$-ball solution~\cite{Coleman1985}.
But, here, we keep the simplest possible theory
with a single three-form gauge field $A$ and a single
corresponding composite  pseudoscalar field $q$.

The dynamics of the $q$-ball solution is
essentially different from Coleman's $Q$-ball~\cite{Coleman1985},
as we do not need the time-dependence of the scalar fields.
The $q$-ball is similar to a
droplet of quantum liquid, because our $q$
(see, in particular, Refs.~\citen{KlinkhamerVolovik2008a,%
KlinkhamerVolovik2008c,KlinkhamerVolovik2016b})
is the analog
of the number density $n$ of the conserved quantity $N$
(particle number). Such a liquid droplet with fixed $N$ has a
fixed radius $R$ and the $q$-ball is similar.

The $q$-ball is expected to be stable for at least three reasons,
two of which will be discussed now and
the third of which will be presented at the end of this section.
First, there are no propagating modes at all outside the $q$-ball,
since the mass of the mode is infinite for $q=0$, as discussed in Sec.~\ref{sec:Propagating-mode}.
Second, the splitting of a $q$-ball into two $q$-balls would lead to
an increase of the surface energy
(for a fixed value of the total charge $Q \equiv q \,V$)
and the splitting is, therefore, excluded.

Expanding on the role of the surface, we note that the
vacuum pressure inside the $q$-ball is determined by
the surface tension $\sigma$ and the radius $R$ of the $q$-ball.
In the absence of gravity, we have (cf. Sec. IV D of Ref.~\citen{KlinkhamerVolovik2008a})
\begin{equation}
P_{\sigma}= - \frac{2\,\sigma}{R}\,,
\label{eq:pressure}
\end{equation}
which corresponds to the effective equation of
state~\cite{KlinkhamerVolovik2008a,Volovik2008}
\begin{equation}
 P_{\sigma}=w_{\sigma}\,\epsilon_{\sigma}\,\,,
 \,\, w_{\sigma}=- \frac{2}{3} \,.
\label{eq:EqState}
\end{equation}

The surface tension $\sigma$ has an order of magnitude
$\epsilon(q_{0})/M$. For a Planck scale $q$-field, we have
the pressure
$P_{\sigma} \sim  -(E_P)^3/R$. The magnitude of $P_{\sigma}$
is of the order of present cosmological constant if $R \sim E_P/H^2$,
which is far beyond the horizon of order $1/H$.

The pressure \eqref{eq:pressure}
from the surface tension is compensated by the bulk vacuum
pressure: $q$ inside the $q$-ball is slightly different from
the perfect-vacuum value $q_0$ (see Sec.~\ref{sec:Numerical-results}
for an explicit numerical result). This is exactly the same
as for a droplet of liquid, where $n$ slightly deviates
from $n_0$ inside the droplet.

Let us, finally, present the third argument for the stability
of the $q$-ball solution (details of the solution
are to follow in Secs.~\ref{sec:Analytical-results} and
\ref{sec:Numerical-results}), where the argument parallels
Coleman's discussion of the standard
$Q$-ball solution~\cite{Coleman1985}.
The $q$-ball can, for large radius $R$,
be considered as a structure with $q\approx q_{0}$ inside
($r<R$) and $q=0$ outside ($r\geq R$).
In the terminology of $Q$-balls, this $q$-ball is similar to the thin-wall $Q$-ball~\cite{Coleman1985}.

Take, as a simple example, the following \textit{Ansatz} for
$\epsilon(q)$:
\begin{equation}
\epsilon(q)= \frac{1}{6}\, q^4 -   \frac{1}{2}\, q_0^2\, q^2   \,,
\label{SimpleAnsatzEpsilon}
\end{equation}
where $q_0$ is again the equilibrium value of the $q$ field,
which is now normalized to have mass dimension 1.
The energy of the $q$-ball with volume $V$ and total charge $Q\equiv q\,V$
is then given by
\begin{equation}
E(V,Q)=V\left( \frac{1}{6}\,\frac{Q^4}{V^4}
- \frac{1}{2}\,  \frac{q_0^2\,Q^2}{V^2}\right)   \,.
\label{EnergyQV}
\end{equation}
Minimization of this energy with respect to the volume $V$
at fixed total charge $Q$ gives
\begin{equation}
\frac{dE}{dV}\bigg |_Q=
\epsilon(q)-q\,\frac{d\epsilon}{d q}
=\frac{1}{2}(q^2_0\,q^2-q^4)= 0 \,.
\label{MinimizationE}
\end{equation}
For a nontrivial $q$-ball, we thus find
the equilibrium value $q=\pm q_0$ for the interior region.
The corresponding equilibrium volume of the $q$-ball equals
\beq\label{MinimizationVolumeV}
V=Q/|q_0|\,.
\eeq
With the charge $Q$ being
conserved~\cite{KlinkhamerVolovik2008b,Aurilia-etal1980,DuncanJensen1989}
(see also Sec.~\ref{subsec:Surface-term})
and the quantity $q_0$ being a ``constant of nature,''
the volume $V$ of the $q$-ball is fixed,
according to \eqref{MinimizationVolumeV}.
With constant volume $V$ of the spherically symmetric $q$-ball,
the edge of the $q$-ball cannot move inwards or outwards,
so that the spherically symmetric $q$-ball solution is stable.

The pressure $P_{V}$ of the vacuum is zero
in the thin-wall approximation,
according to the Gibbs--Duhem relation
$\epsilon - q\, d\epsilon/dq = - P_{V}$
and the result \eqref{MinimizationE}.
A nonzero value for the pressure comes from
the surface term \eqref{eq:pressure}.
To summarize, the $q$-ball is stable due to the conservation of the
total charge $Q$, while the surface term is only responsible for the
pressure inside the $q$-ball.

\section{Analytical results}
\label{sec:Analytical-results}

The profile of the thin wall separating the true vacuum and
the absolutely empty vacuum can be found analytically
in the limit of large $R$, where the wall can be considered as a flat plane and the profile depends on the spatial coordinate $x$ across the wall with
$q=0$ at $-\infty<x\leq 0$ and  $q^2>0$ at $0<x<\infty$.  The thickness of the transition region is $d\sim M^{-1}$,
with $M^2$ given by \eqref{eq:M-square}.

\subsection{Surface term}
\label{subsec:Surface-term}

In order to find this $q(x)$ profile, we need to establish the
relevant equation.
Consider the function $q(x)$ at $x>0$ with the  boundary conditions
$q(0)=0$ and $q(\infty)=q_{0}$.
For $x>0$ in flat spacetime, the generalized Maxwell equation \eqref{eq:Maxwell} reads
\begin{equation}\label{eq:Maxwellx1}
\partial_x \left(
\frac{d\epsilon(q)}{d q}+
\frac{1}{8}\, \frac{dK(q)}{dq}\, \Big[\partial_x  (q^2)\Big]^2
-
\frac{1}{2}\,
q\, \partial_x  \Big[K(q)\,\partial_x(q^2)\Big]
\right) =0
\end{equation}
and has the solution
\begin{equation}\label{eq:Maxwellx2}
\frac{d\epsilon(q)}{d q}+
\frac{1}{8}\, \frac{dK(q)}{dq}\, \Big[\partial_x  (q^2)\Big]^2
-
\frac{1}{2}\,
q\, \partial_x  \Big[K(q)\,\partial_x (q^2)\Big] =\mu\,,
\end{equation}
with integration constant (chemical potential) $\mu$.
In the limit of infinite radius $R$, the chemical potential $\mu$
is the same as for the infinite-volume equilibrium vacuum
and we have $\mu = \mu_{0}$ in \eqref{eq:Maxwellx2} for $x>0$.
For $x<0$, we have $q=0$ and thus $\mu=0$ in \eqref{eq:Maxwellx2}.
In short, the $q$-ball dynamics follows from a single
ordinary differential equation (ODE),
given by \eqref{eq:Maxwellx2} with
\beq\label{eq:Maxwellx2-mu-jump}
\mu
= \left\{  \begin{array}{lcl}
             0       &\;\;\text{for}\;\;& x \leq 0   \,, \\
             \mu_0   &\;\;\text{for}\;\;& x >    0 \,.
           \end{array}
           \right.
\eeq

This jump in the chemical
potential at $x=0$  (from $\mu=0$ for negative $x$ values
to $\mu=\mu_0$ for positive $x$ values) corresponds to a
delta-function source term $\mu_0\,\delta(x)$
on the right-hand side of \eqref{eq:Maxwellx1}. Such a source term may be obtained by the introduction of a surface
contribution in the action \eqref{eq:actionF},
 \begin{equation}
 S_{\text{surface}-\mathbb{S}}= -\mu_\mathbb{S}\, \int_{\mathbb{S}}
\,dS^{\alpha\beta\gamma} A_{\alpha\beta\gamma}  \,,
\label{eq:surfaceaction}
\end{equation}
for a boundaryless 3-dimensional surface $\mathbb{S}$
and a constant $\mu_\mathbb{S}$.
This  surface term describes the conservation of the flux of the four-form
$F$-field. Since
 \begin{equation}
\int_{\mathbb{S}\,=\,\partial \mathbb{V}}\,
dS^{\alpha\beta\gamma} \, A_{\alpha\beta\gamma}
= \int_{\mathbb{V}}\,
dS^{\alpha\beta\gamma\delta}\, F_{\alpha\beta\gamma\delta} \,,
\label{eq:surfaceTo4volume}
\end{equation}
we obtain, for the time-independent $q$-field of the infinite-radius $q$-ball and with $\mu_\mathbb{S}=\mu_0$, a term corresponding to the conservation of the total charge $Q$ of the $q$-ball:
 \begin{equation}
 S_{\text{surface}-\mathbb{S}}^\text{($q$-ball,\;$R=\infty$)}=
-\mu_0 \, \int dt\, \int d^3x \: q =-\mu_0\,\int dt \: Q  \,.
\label{eq:timeslice}
\end{equation}

For the flat $x=0$ surface between empty and equilibrium vacua,
the variation of \eqref{eq:surfaceaction} with $\mu_\mathbb{S}=\mu_0$
over the three-form gauge field $A$ gives the a term $\mu_0\, \delta(x)$
on the right-hand side of \eqref{eq:Maxwellx1},
so that we get for the infinite-radius $q$-ball
\begin{equation}\label{eq:Maxwellx3}
\partial_x \left(
\frac{d\epsilon(q)}{d q}+
\frac{1}{8}\, \frac{dK(q)}{dq}\, \Big[\partial_x  (q^2)\Big]^2
-
\frac{1}{2}\,
q\, \partial_x  \,\Big[K(q)\,\partial_x (q^2)\Big]
\right) =
\mu_0 \,\delta(x)\,,
\end{equation}
with the empty-vacuum boundary condition $\lim_{x\to -\infty} q(x)=0$.

The previous discussion was specialized to the case of the
infinite-radius $q$-ball, even though the
surface term \eqref{eq:surfaceaction} was given in
its most general form.
The same discussion holds for the finite-radius $q$-ball
with the coordinate $x$ replaced by the radial coordinate $r$
and the constant $\mu_\mathbb{S}$ taking the appropriate value $\mu_{c}$
in order to cancel the surface pressure \eqref{eq:pressure};
see Sec.~\ref{sec:Numerical-results} for details.

To summarize, the surface term \eqref{eq:surfaceaction}
gives precisely the delta-function source term in the
generalized Maxwell equation.
It properly reflects the conservation law for
the charge $Q \equiv \int_{V} d^{3}x \, q(x)$
of the $q$-ball solution surrounded
by the absolutely empty vacuum as discussed in Sec.~\ref{sec:Kinetic-term}.

\subsection{Profile of the infinite-radius $q$-ball }
\label{subsec:Profile-infinite-radiusq-ball}

Let us now determine the detailed behavior of the $q$-field
on the true-vacuum side at $x>0$.
For the theory \eqref{eq:actionF-Fdefinition12},
the function $q(x)$ is the extremum of the following thermodynamic potential which can be interpreted as the surface tension of the $q$-ball:
\begin{equation}
\sigma^{(R\to\infty)}=\int_{0}^{\infty} \,dx\,\left(\epsilon(q)-\mu_{0}\, q
+\frac{1}{8}\,K(q)\, \partial_x (q^2)\, \partial_x(q^2) \right) \,.
\label{eq:1D}
\end{equation}
In the limit of infinite radius $R$,
the chemical potential on the positive $x$ side is the same as
for the infinite-volume equilibrium vacuum,
which is why we use $\mu = \mu_{0}$ in \eqref{eq:1D}.
In Sec.~\ref{sec:Numerical-results}, we will see
that the finite-radius $q$-ball has $\mu \ne \mu_{0}$.

The solution of the second-order equation for $q$
[found by variation of \eqref{eq:1D}]
can be obtained from the first integral, which gives
\begin{equation}
\epsilon(q)-\mu_{0}\, q
-\frac{1}{8}\,K(q)\, \partial_x (q^2) \,\partial_x(q^2) ={\rm const}=0\,.
\label{eq:FirstIntegral}
\end{equation}
From the above equation we get
\begin{equation}
\int_{0}^x dx' = \int_{0}^{q^2<q_{0}^2} d(q^{\prime\,2})\,\sqrt{\frac{K(q')}{8\,[\epsilon(q')-\mu_{0}\, q']}  }\,.
\label{eq:FirstIntegral2}
\end{equation}
For $x\rightarrow \infty$,  $q^2$ approaches $q_{0}^2$ from below.
For given functions of $\epsilon(q)$ and $K(q)$,
we can always find the solution of \eqref{eq:FirstIntegral2} numerically.
Still,
keeping $\epsilon(q)$ and $K(q)$ generic, the asymptotic behavior for
$x\rightarrow \infty$ and $x\rightarrow 0$ can be found analytically.

The integral over $q^2$ is concentrated near $q_{0}^2$, where it diverges
logarithmically. In terms of the variable $\varphi$ from
\eqref{eq:q-field-scalar-mode}, we obtain the following
asymptotic behavior as $x\rightarrow \infty$:
\begin{equation}
x \rightarrow  \int^{1}_{\varphi}  \frac{d\varphi'}{M\, \varphi'} + \text{const} =
- \frac{1}{M}\,  \ln \varphi + \text{const}\,,
\label{eq:asymptote-infinity-a}
\end{equation}
or
\begin{equation}
\varphi(x)\propto e^{-Mx} \;\;\text{for}\;\;x\rightarrow \infty \,.
\label{eq:asymptote-infinity-b}
\end{equation}

The asymptote at $x\rightarrow 0$ is obtained from expansion near $q=0$.
If $K(0)$ is positive and finite, we have
\begin{equation}
x = \int_{0}^q \,q'dq'\,\sqrt{\frac{K(0)}{2\,q'\,|\mu_{0}|} } = \frac{2}{3}\,\sqrt{\frac{K(0)}{\,2|\mu_{0}|} }\,q^{3/2}\,,
\label{eq:asymptote-zero-a}
\end{equation}
which gives
\beq
q(x)\propto x^{2/3} \;\;\text{for}\;\;x\rightarrow 0\,.
\label{eq:asymptote-zero-b}
\eeq
If $K(q) \propto q^{\beta}$ as $q\rightarrow 0$ with $\beta>0$,
we have $q(x)\propto x^{2/(3+\beta)}$ as $x\rightarrow 0$.

It is also possible to obtain the complete $q$-ball profile
analytically for certain special \textit{Ans\"{a}tze}
of the functions $K(q)$ and $\epsilon(q)$.
Let us give an example. We change to dimensionless variables
($x$ giving $s$,  $q$ giving $f$, and $\mu_0$ giving $u_0$)
and make the following \textit{Ans\"{a}tze}
for the dimensionless versions of the coupling $K(q)$ of the
$q$ kinetic term, the gravitational coupling $G(q)$, and
the energy density $\epsilon(q)$
[cf. Eq.~(\ref{SimpleAnsatzEpsilon}) above]:
\bsubeqs\label{eq:Ansaetze}
\beqa
\label{eq:Ansaetze-k}
k(f) &=& 1 \,,
\\[2mm]
\label{eq:Ansaetze-g}
g(f) &=& 0 \,,
\\[2mm]
\label{eq:Ansaetze-epsilon}
\epsilon(f) &=& \frac{1}{2}\,f^2\,\left( \frac{1}{3}\,f^2-1\right) \,,
\eeqa
which give the following dimensionless constants:
\beqa
\label{eq:Ansaetze-f0}
f_0  &=& 1 \,,
\\[2mm]
\label{eq:Ansaetze-u0}
u_0  &=& -\frac{1}{3} \,,
\\[2mm]
\label{eq:Ansaetze-m2}
m^2  &=& 1\,,
\eeqa
\esubeqs
with $m^2$ corresponding to the dimensionless version of
\eqref{eq:M-square}.
Note that precisely this \textit{Ansatz} for $\epsilon(f)$
has been used in Ref.~\citen{KlinkhamerVolovik2008b}
and that we set the dimensionless gravitational coupling $g$
to zero, so that Minkowski spacetime can be used consistently.
For later use, we also give the dimensionless version
of the integrand of \eqref{eq:1D} for the \textit{Ans\"{a}tze} \eqref{eq:Ansaetze},
\beq\label{eq:widetilde-e}
r_{\sigma}\,[f]=
\epsilon(f) - u_0\,f + \half\, f^2\, (\partial_s f)^2\,,
\eeq
with $\epsilon(f)$ from \eqref{eq:Ansaetze-epsilon}
and $u_0$ from \eqref{eq:Ansaetze-u0}.

\begin{figure*}[t]  
\vspace*{-0cm}      
\begin{center}
\includegraphics[width=0.75\textwidth]{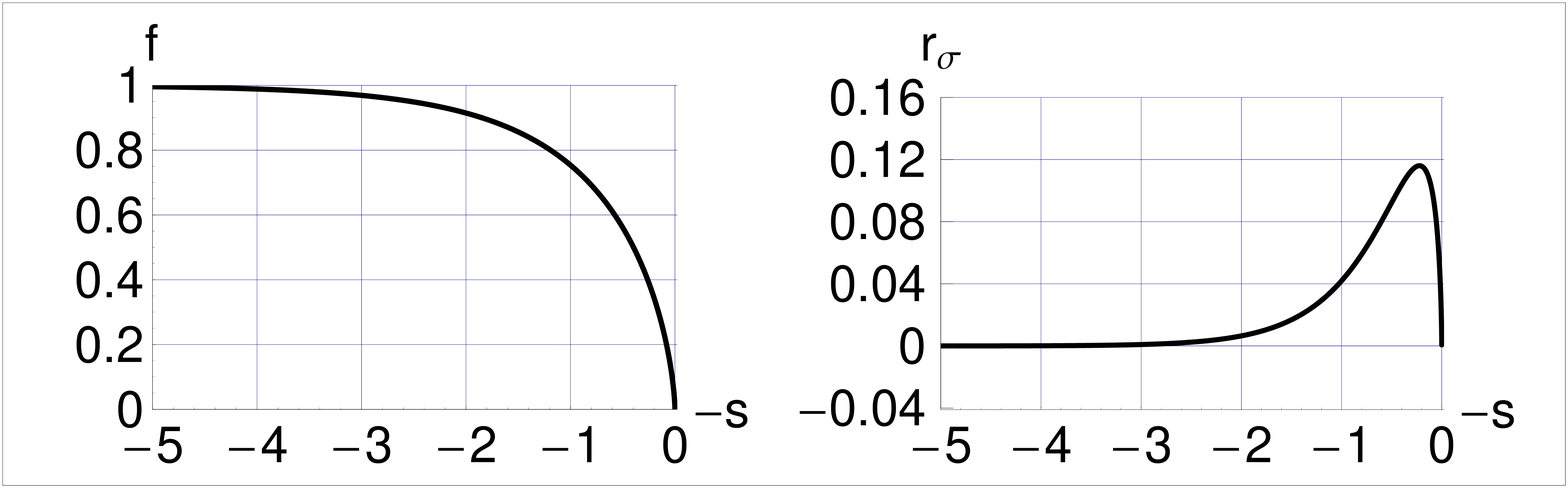}
\end{center}
\vspace*{-0mm}
\caption{The left panel gives the analytic infinite-radius $q$-ball
profile $f(s)$ from \eqref{eq:analytic-q-ball-profile}, with
$f(0)=0$ and $f\uparrow f_0=1$ for $-s\to-\infty$.
Here, $f$ refers to the dimensionless $q$-field and $s$ to
the dimensionless spatial coordinate orthogonal to the edge of the
infinite-radius $q$-ball. The right panel gives
the corresponding energy density \eqref{eq:widetilde-e}.
}
\label{fig:analytic-q-ball-profile}
\end{figure*}

The dimensionless version of \eqref{eq:FirstIntegral2}
for the \textit{Ans\"{a}tze} \eqref{eq:Ansaetze} then has
the following analytic solution:
\beqa\label{eq:analytic-q-ball-profile}
s &=&
\frac{2}{{\sqrt{3}}}\,\left( 1 - f \right) \,
{\sqrt{\frac{2 + f}{2\, - 3\,f + f^3}}}\,
\nonumber\\[1mm]
&&
\times
    \left(
      {\sqrt{3}}\:\text{arctanh}
      \left({\sqrt{\frac{3\,f}{2 + f}}}\,\right)
      - 3\:\text{arcsinh}
      \left(\sqrt{\frac{f}{2}}\,\right)
      \right) \,.
\eeqa
The asymptotic behavior of
the solution \eqref{eq:analytic-q-ball-profile}
agrees with the previous general
results \eqref{eq:asymptote-infinity-b} and \eqref{eq:asymptote-zero-b}.
Equation \eqref{eq:analytic-q-ball-profile} defines the
function $s(f)$ and a plot of the inverse function $f(s)$
is shown in Fig.~\ref{fig:analytic-q-ball-profile},
with reflected $s$ values for later comparison.
Figure~\ref{fig:analytic-q-ball-profile} also shows the
energy density \eqref{eq:widetilde-e}.

\section{Numerical results}
\label{sec:Numerical-results}

For the finite-radius $q$-ball in Minkowski spacetime,
the generalized Maxwell equation \eqref{eq:Maxwell}
for $K(q)=K=\text{const}$ reads in spherical coordinates
\beq\label{eq:Maxwell-q-ball}
\partial_r \left(
\frac{d\epsilon(q)}{d q}
-
\frac{1}{2}\, K\,
q\,\frac{1}{r}\, \partial_r\partial_r \Big[r\,q^2\Big]\right) =0\,,
\eeq
with the following boundary conditions on $q(r)$:
\bsubeqs\label{eq:Maxwell-q-ball-bcs}
\beqa\label{eq:Maxwell-q-ball-bc-q0}
q(0)&=&q_{c}  > 0\,,
\\[2mm]
\label{eq:Maxwell-q-ball-bc-qprime0}
q'(0)&=&0\,,
\\[2mm]
\label{eq:Maxwell-q-ball-bc-qR}
q(R)&=&0\,.
\eeqa
\esubeqs
Note that the two boundary conditions \eqref{eq:Maxwell-q-ball-bc-q0}
and \eqref{eq:Maxwell-q-ball-bc-qR} exclude having a constant solution.

As discussed in Sec.~\ref{subsec:Surface-term},
the $q$-ball requires that the right-hand side
of \eqref{eq:Maxwell-q-ball} be replaced by a delta-function.
In order to obtain the numerical solution $q(r)$, we proceed as follows.
We take the \textit{Ans\"{a}tze} \eqref{eq:Ansaetze}
and use the same dimensionless variables as in
Sec.~\ref{subsec:Profile-infinite-radiusq-ball},
together with the dimensionless radial coordinate $v \in [0,\,\infty)$.
Next, we replace the third-order ODE \eqref{eq:Maxwell-q-ball}
by two second-order ODEs, one for the inside region and
one for the outside region,
\beq\label{eq:Maxwell-q-ball-2nd-order}
\frac{d\epsilon(f)}{d f}-
\frac{1}{2}\,
f\,\frac{1}{v}\, \partial_v\partial_v \Big[v\, f^2\Big]
= \left\{  \begin{array}{lcl}
             u_{c} &\;\;\text{for}\;\;& 0 \leq v < v_R   \,, \\
             0   &\;\;\text{for}\;\;& v \geq v_R \,,
           \end{array}
 \right.
\eeq
with $v_R>0$ and $\epsilon(f)$ from \eqref{eq:Ansaetze-epsilon}.
The following boundary conditions on $f(v)$ hold for the
inside region:
\bsubeqs\label{eq:q-ball-solution-inside-outside-bcs}
\beq\label{eq:q-ball-solution-inside-bcs}
f(0)=f_{c} >0\,,\;\;
\lim_{v\uparrow v_R}\;f(v)=0 \,,
\eeq
and for the outside region:
\beq
\label{eq:q-ball-solution-outside-bcs}
f(v_R)=0\,, \;\;f'(v_R)=0 \,.
\eeq
\esubeqs
For later reference, we also give the dimensionless interior
energy density functional,
\beq\label{eq:energy-density}
r_{c}[f] \equiv \epsilon(f) - u_{c}\,f + \half\, f^2\, (\partial_v f)^2\,,
\eeq
with $\epsilon(f)$ from \eqref{eq:Ansaetze-epsilon}
and $u_{c}$ from \eqref{eq:Maxwell-q-ball-2nd-order}.

The ODE \eqref{eq:Maxwell-q-ball-2nd-order} with
boundary conditions \eqref{eq:q-ball-solution-inside-outside-bcs}
gives the $q$-ball solution
with a sharp drop of the integration constant $u$ at the $q$-ball edge
and the corresponding non-smoothness of the function $f(v)$.
In line with the discussion
of Sec.~\ref{subsec:Surface-term},
we observe that the solution of \eqref{eq:Maxwell-q-ball-2nd-order}
solves \eqref{eq:Maxwell-q-ball} for $r \ne R$ (or $v \ne v_R$)
but \underline{not} at $r = R$ (corresponding to the sphere $S_R$).
The proper solution
is determined by the surface term \eqref{eq:surfaceaction}, which describes the interface  between the physical vacuum and the
absolutely empty vacuum.

The ODE \eqref{eq:Maxwell-q-ball-2nd-order}
for the outside region with boundary conditions
\eqref{eq:q-ball-solution-outside-bcs} gives immediately the solution
\beq\label{eq:q-ball-solution-outside-solution}
f(v)=0 \;\;\;\text{for}\;\; v \geq v_R\,.
\eeq
But the ODE for the inside region is singular and requires some care.
We solve the ODE \eqref{eq:Maxwell-q-ball-2nd-order}
over the $v$ interval $\{v_\text{start},\,v_\text{end}\}$
with $0<v_\text{start}<v_\text{end}$ and take boundary conditions
$\{f(v_\text{start}),\,f'(v_\text{start})\}=\{f_{c},\, 0\}$
with $f_{c}>0$.
By choosing appropriate values for $f_{c}$ and $u_{c}$,
we obtain a solution with $f(v_\text{end})=0$,
so that we can identify $v_\text{end}$ as the dimensionless
parameter $v_R$ corresponding to the radius $R$ of the
$q$-ball.

\begin{figure*}[t]  
\vspace*{-0cm}      
\begin{center}      
\includegraphics[width=0.75\textwidth]{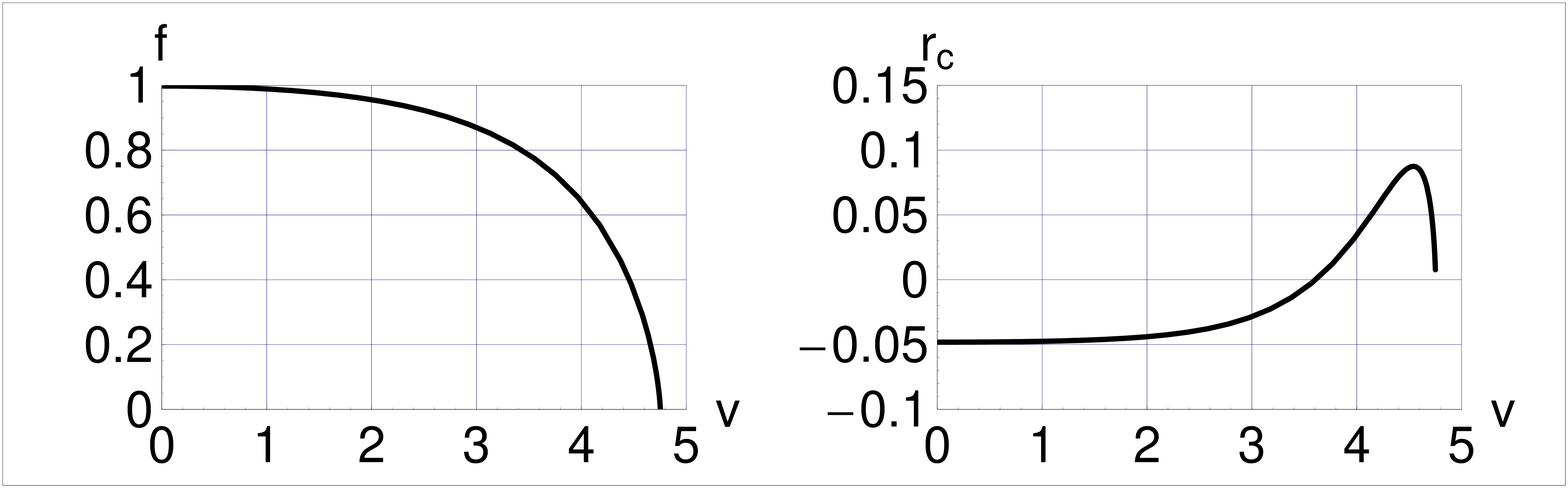}
\end{center}
\vspace*{-0mm}
\caption{The left panel gives the numeric finite-radius $q$-ball
profile $f(v)$
from the second-order ODE \eqref{eq:Maxwell-q-ball-2nd-order}
with parameters and boundary conditions \eqref{eq:numerical-parameters}.
Here, $f$ refers to the dimensionless $q$-field and $v$ to
the dimensionless radial coordinate of the $q$-ball.
The right panel gives the corresponding energy density   \eqref{eq:energy-density}.
}
\label{fig:numeric-q-ball-profile}
\end{figure*}

Specifically, we take the values
\bsubeqs\label{eq:numerical-parameters}
\beqa\label{eq:numerical-parameters-range}
\{v_\text{start},\,v_\text{end}\}&=&
\{10^{-2},\, 4.74\}\,,
\\[2mm]
\label{eq:numerical-parameters-bcs}
\{f(v_\text{start}),\,f'(v_\text{start}),\,u_{c}\}&=&
\{0.998,\, 0,\, -0.285\}\,.
\eeqa
\esubeqs
Figure~\ref{fig:numeric-q-ball-profile} shows the
resulting numerical solution
$f(v)$ for $v< v_R \approx 4.74$
(similar results have been obtained with
$v_\text{start}=10^{-3}$, the other parameters and
boundary conditions being kept the same).
As mentioned above, we have $f(v)=0$ for $v> v_R$.
The energy density $r_{c}$  in Fig.~\ref{fig:numeric-q-ball-profile}
is seen to remain finite for $v< v_R$ and to approach
(within numerical errors) a zero value for $v\uparrow v_R$
[the exact solution \eqref{eq:q-ball-solution-outside-solution}
has $r[f] \equiv\epsilon(f) + \half\, f^2\, (\partial_v f)^2=0$ for $v\geq v_R$\,].
Note that, especially near the edge at $f=0$,
the $f$-profile of the finite-radius $q$-ball
in Fig.~\ref{fig:numeric-q-ball-profile} is close
to that of the infinite-radius $q$-ball
in Fig.~\ref{fig:analytic-q-ball-profile}.

\section{Outlook}
\label{Outlook}

We have used, in the present article,
a $q$-field realized by a four-form field strength $F$
as a phenomenological description of the deep vacuum.
But, there are also other types of $q$-fields,
provided they obey
the condition necessary to represent a stable quantum vacuum.
This necessary condition is the
conservation law for the vacuum variable $q$,
which is represented  by \eqref{eq:Maxwell} here.
Note that a fundamental scalar field does not obey this condition.
The structure of \eqref{eq:Maxwell}
holds also for the other types of vacuum $q$-fields.
This demonstrates the universality of the phenomenology of the
quantum vacuum:  the phenomenology does not really depend
 on the details at the microscopic  scale
(be it Planckian or trans-Planckian) but is determined by
phenomenological functions of the vacuum variable $q$,
such as the energy density $\epsilon(q)$,
the gravitational coupling $G(q)$, and the rigidity $K(q)$.
These functions determine, in particular,
the compressibility $\chi(q)$ of the quantum vacuum,
the vacuum pressure $P(q)$, and the mass $M(q)$ of the propagating mode.

This universal behavior of $q$-theory is similar to
the universal description of stable
quantum many-body systems, such as quantum liquids formulated
in terms of the number density $n$ of atoms
(the nonrelativistic counterpart of our vacuum variable $q$).
The vacuum $q$-ball considered in the present article is, in fact, the
analog of a droplet of quantum liquid, which has a stable ground state
at nonzero $n_{c}$ inside the droplet and $n=0$ outside, with a radius $R$
determined by the total number $N$ of atoms in the droplet.
The chemical potential $\mu$
(being conjugate to $n$ in thermodynamics and acting as a Lagrange
multiplier in the Hamiltonian)
adjusts in order to match the external pressure in equilibrium.
Without external forces on the droplet, the
internal pressure is determined by the surface tension $\sigma$.
In the limit of a large droplet radius, $R\rightarrow \infty$,
the thermodynamic potential
$\epsilon(n) - \mu\, n$ is nullified:
$\epsilon(n) - \mu\, n=-P = P_{\sigma} \rightarrow 0$
and $\mu \rightarrow \mu_0$.
This thermodynamic potential in terms of the atom number
density $n$ is the analog of the
gravitating vacuum energy density (cosmological constant)
in terms of the vacuum variable $q$.

The space outside a droplet of quantum liquid,
being free of atoms,  is analogous to the
absolutely empty vacuum of our discussion.
The latter is a new concept and corresponds to a vacuum
without any field and corresponding quantum fluctuations.
The $q$-ball considered in the present article
is a droplet of physical vacuum inside this absolutely empty vacuum. The extended $q$-theory with gradient terms allows us to describe the structure of the boundary of the $q$-ball.
The $q$-ball is stable due to the conservation of the
total charge $Q$, while the surface term is only responsible for the
pressure inside the $q$-ball.

In the above discussion,
it has been assumed that the physical vacuum is in full
equilibrium and that its chemical potential $\mu$ has the equilibrium
value $\mu_0$. The next step is to study the dynamics of the $q$-ball,
which describes the relaxation to equilibrium
from an arbitrary initial state of the $q$-ball.
We expect a similar behavior as
for the droplet of quantum liquid: nonequilibrium processes
take place, accompanied by oscillations of the
droplet shape and by radiation of the surplus droplet energy,
and the equilibrium state of the liquid droplet is reached
in the end. In this equilibration process,
the chemical potential is no longer a
Lagrange multiplier but becomes a time-dependent variable, which
relaxes to its equilibrium value.

It will be interesting to study the
process of equilibration of the $q$-field, both in the absence and in the presence of gravity.
The case with gravity may give a hint of how
our Universe has relaxed to its present state,
which is extremely close to equilibrium.

\vspace*{-3mm}
\section*{Note Added}

After completion of the present article,
a further potential manifestation of the $q$-field with
effective action \eqref{eq:actionF} has been found,
namely, as a possible
description of cold dark matter~\cite{KlinkhamerVolovik2016c}.
In addition, it was shown that
the higher-derivative $q$-theory \eqref{eq:actionF-Fdefinition12}
does not suffer from the Ostrogradsky instability at the classical
level.~\cite{KlinkhamerMistele2017}

\vspace*{-3mm}
\section*{Acknowledgments}

We thank T. Mistele and
M. Savelainen for correcting an error in
an earlier version of Eq.~\eqref{eq:Maxwell}.
The work of GEV has been supported by the European Research Council
(ERC) under the European Union's Horizon 2020 research and innovation
programme (Grant Agreement No. 694248).



\end{document}